\documentclass[prb,showpacs,a4paper,floatfix,twocolumn]{revtex4}
\usepackage{graphicx}
\usepackage{amssymb,amsfonts,amsmath}
\usepackage[T1]{fontenc}
\usepackage{times}
\usepackage{color}

\begin{document}
\title{Finite-Temperature N\'eel Ordering of Fluctuations in a
  Plaquette Orbital Model} \date{\today} \author{Sandro
  \surname{Wenzel}}\altaffiliation[Present
  address: ]{Max-Planck-Institute for Physics of Complex Systems,
  N\"othnitzer Str.\ 38, 01187 Dresden, Germany}\email{wenzel@itp.uni-leipzig.de}
\affiliation{Institut f\"ur Theoretische Physik and Centre for
  Theoretical Sciences (NTZ), Universit\"at Leipzig, Postfach
  $100\,920$, D-$04009$ Leipzig, Germany}
\author{Wolfhard  \surname{Janke}} \email{janke@itp.uni-leipzig.de}
\affiliation{Institut f\"ur Theoretische Physik and Centre for
  Theoretical Sciences (NTZ), Universit\"at Leipzig, Postfach
  $100\,920$, D-$04009$ Leipzig, Germany}

\begin{abstract}
  We present a pseudo-spin model which should be experimentally
  accessible using solid-state devices and, being a variation on the
  compass model, adds to the toolbox for the protection of qubits in the area of
  quantum information. Using Monte Carlo methods, we find for both
  classical and quantum spins in two and three dimensions Ising type
  N\'eel ordering of energy fluctuations at finite temperatures
  without magnetic order. We also readdress the controversy concerning
  the stability of the ordered state in the presence of quenched
  impurities and present numerical results which are at clear variance
  with earlier claims in the literature.
\end{abstract}

\maketitle

\section{Introduction}
The prospect of topological quantum computation to implement fault
tolerant quantum bits has led to considerable interest in the field
\cite{kitaev-toriccode} over the past years. A particular route within
this area is the construction of simple (spin) models, as microscopic
models of topological field theories, that allow a direct experimental
realization.  A hallmark is the so-called Kitaev model,\cite{kitaev-model} which is exactly solvable and known to be
implementable as well as controllable using optical devices.\cite{experiments} 
Very recently, related efforts have been spent to construct similar models which can be realized using solid-state techniques.\cite{GladchenkJackeli} Guided by a few principles -- a degenerate
ground-state manifold and a gap to excited states -- the so-called
compass model (CM), with the Hamiltonian
\begin{equation}
\label{eq:hamiltonian0}
\mathcal{H}_\mathrm{CM}=J_x\sum_i  S^x_i S^x_{i+e_x} + J_z \sum_i S^z_i S^z_{i+e_z}\,,
\end{equation}
on an $L\times L$ lattice, was proposed as a simple model allowing for the protection of qubits.\cite{doucot:024505} The CM is realizable by a particular
arrangement of Josephson-junction devices and can be described by a
$Z_2$ Chern-Simons topological quantum field theory.\cite{doucot:024505} Extensions of \eqref{eq:hamiltonian0} to global
interactions possess even better fault tolerant properties.\cite{milman:020503}

Originally introduced as an orbital model for Mott
insulators,\cite{kugel82} research into the actual CM has been pushed by several
groups, %
which established degenerate ground-state
properties,\cite{dorier:024448} first-order quantum phase
transitions,\cite{firstorder} relation to $p+ip$
superconductivity,\cite{nussinovpip} and the existence of directional
order.\cite{mishra:207201,tanaka:256402,wenzelQCMPRB} Reference
\onlinecite{tanaka:256402} argues that quantum spins support a
resistivity of the ordered phase towards quenched disorder, which is
in sharp contrast to classical degrees of freedom for which the
ordered phase vanishes rapidly with increasing disorder. By a detailed
Monte Carlo (MC) study\cite{wenzelQCMPRB} of the quantum and classical
CM, we have recently shown that this conclusion, however, needs
further support as the CM shows unusual and extremely slowly converging
finite-size scaling (FSS) properties on periodic lattices, which were
used in Ref.~\onlinecite{tanaka:256402}.  Recently, an interesting
extension of \eqref{eq:hamiltonian0} to include a magnetic field term
$hS$ was performed, leading to \emph{thermal canting of spin
  order}.\cite{scarola:CM}

In the search for other fundamental spin models and to gain further
insights into the field around the Kitaev model, we propose here a
different -- \emph{geometric} -- modification of the CM and
concisely report on its intriguing physics. Our main
result is the establishment of an interesting ordering that
can be described by a crystallization and modulation of local energy
contributions but which lacks conventional magnetic order. We show
that the proposed model falls into the Ising universality class and that it
possesses well behaved FSS properties in contrast to the CM. 
%
In the last part of this paper, we use this advantage to
investigate the influence of (weak) quenched disorder in form of
random vacancies to study their influence on the nature of the phase
transition. We show that long-range order is completely lost already for
very weak impurity concentrations.
%

\section{The model}
\begin{figure}[t]
\includegraphics[width=0.65\columnwidth]{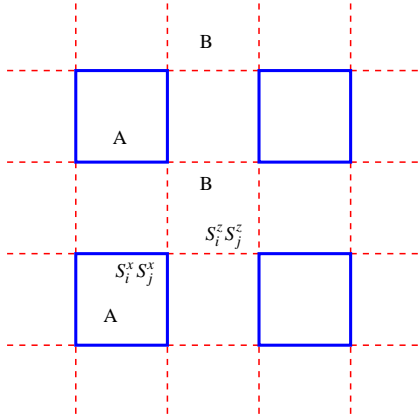}
\caption{\label{fig:lattice} (color online). Illustration of the POM
  lattice.  The blue (thick) bonds indicate $S^xS^x$ terms while the red
  (dashed) bonds stand for $S^zS^z$ links.  The lattice is closely
  related to a checkerboard. The generalization to the
  3D POM is straightforward with sub-lattices A and B
  being cubes rather than plaquettes.}
\end{figure}

The plaquette orbital model (POM) is defined by the Hamiltonian
\begin{equation}
\label{eq:hamiltonian1}
\mathcal{H}_\mathrm{POM}=J_A\sum_{\langle i,j \rangle_A} S^x_i S^x_j + J_B \sum_{\langle i,j \rangle_B} S^z_i S^z_j\,,
\end{equation}
where $S^x$ and $S^z$ are components of a two-component spin $S$,
which can represent both classical and quantum degrees of freedom. In
the latter case $S^x$ and $S^z$ are represented by the usual $S=1/2$
Pauli matrices while in the classical case they denote projections of
a continuous spin parameterized by an angle $\theta$ on the unit
sphere.  The bonds $\langle i,j \rangle_A$, $\langle i,j \rangle_B$ on
sub-lattices A and B are arranged as depicted in
Fig.~\ref{fig:lattice}. The coupling strengths $J_A$ and $J_B$ are in
principle arbitrary. Here, we are interested in the isotropic case
$J_A=J_B=J=1$. The sign of $J$ has no relevance since it can be transformed away on
bipartite lattices.\cite{mishra:207201} With $N=L^d$ we denote the
number of spins on a cubic lattice of linear extension $L$ and
dimension $d$. It should be emphasized that in contrast to the CM in three dimensions
(3D) or the Kitaev model in two dimensions (2D), quantum MC
investigations of the POM can easily be done also in 3D since there is
no sign problem.

Note that the Hamiltonian \eqref{eq:hamiltonian1} is $Z_2$ symmetric
under exchange of sub-lattices A and B and spin indices $x$ and $z$.
Define further four-spin operators \begin{equation}P_r = S_r^z S_{r+e_x}^z S_{r+e_y}^z
S_{r+e_x+e_y}^z\end{equation} and \begin{equation}Q_l = S_l^x S_{l+e_x}^x S_{l+e_y}^x
S_{l+e_y+e_x}^x,\end{equation} with $e_x$ and $e_y$ being unit vectors on the
lattice and $r$ ($l$) a site pointing to the lower left corner of A
(B) plaquettes. We can show that
$[\mathcal{H}_\mathrm{{POM}},P_r]=[\mathcal{H}_\mathrm{{POM}},Q_l]=0$
are local symmetries. Hence, on A plaquettes the operations $S^x_i \to -
S^x_i\;, S_i^z \to S_i^z$ are symmetries with the analog expression on B
plaquettes reading $S_i^x\to S_i^x\;, S_i^z \to -S_i^z$.\cite{dorier:024448}  For any $r$ and
$L>2$, there further exists at least one index $l$ such that
$[P_r,Q_l]\neq 0$.\cite{footnote1} This shows that every energy eigenvalue of
$\mathcal{H}_\mathrm{POM}$ is at least two-fold degenerate. Performing
an exact diagonalization employing invariant subspaces of the operator
$P$ (see Ref. \onlinecite{dorier:024448}), we could obtain all
eigenvalues on a $N=4\times 4$ cluster confirming this conclusion.
The POM hence possesses the same behavior as the CM in this regard.
Whether the excitation gap persists in the infinite-volume limit remains to be investigated. The 3D extension of the model is obvious. Every plaquette
becomes a cube, otherwise all arguments stay the same.

We now turn to a stochastic investigation of the model using
established MC methods ranging from Metropolis sampling for classical
variables to the quantum stochastic series expansion
(SSE).\cite{PhysRevB.59.R14157,PhysRevE.66.046701} Key to succesfully
simulate model \eqref{eq:hamiltonian1} on large lattices is the use of
parallel tempering methods to avoid barriers and reduce
autocorrelation times. Details of our approach can be found in
Ref.~\onlinecite{wenzelQCMPRB}.

\section{Results}
\subsection{N\'eel ordering}
The exchange symmetry ($A\Leftrightarrow B$) provides the possibility  for spontaneous symmetry breaking.
\begin{figure}
\includegraphics[width=0.96\columnwidth]{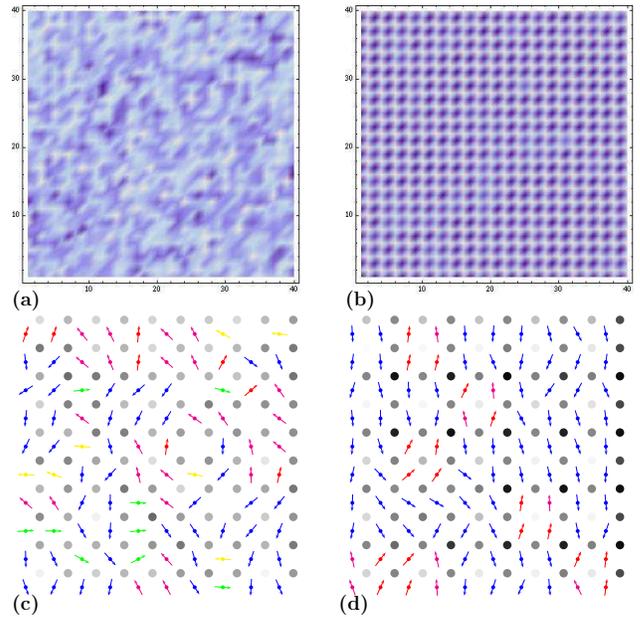}
\caption{\label{fig:neelstate} (color online). Snapshots of the
  plaquette energy distribution on a $N=40\times 40$ lattice taken at
  $T=1.0$ and $T=0.02$, respectively. In the high-temperature regime (a)
  the system is disordered and a symmetry breaking has occurred in (b)
  for $T$ less than a critical temperature $T_\mathrm{c}$. Darker regions
  (color) represent lower energy. Figures (c)
  and (d) are snapshots of configurations in spin space corresponding to the high- and
  low-temperature phases for $L=10$. No evident magnetic order is seen. The
  circles in (c) and (d) signify local energy density where darker
  means larger negative energy. Spins are color (gray) coded to make their
  direction more apparent.}
\end{figure}
To see if this exchange symmetry is broken for some temperature
$T<T_\mathrm{c}$ a suitable order parameter should be defined. Let us
just consider energy fluctuation on the two sub-lattices. Then,
following Refs.~\onlinecite{tanaka:256402,Batista_dimred2005,wenzelQCMPRB} we define the quantity
\begin{equation}
\label{eq:orderD}
D=\frac{1}{N} | E_A- E_B |\,,
\end{equation}
where $E_A = J_A \sum_{\langle i,j\rangle_A} S_i^x S_j^x$ is the
energy contribution on sub-lattice A with the obvious relation for
$E_B$. The suitability of that quantity can be seen in
Fig.~\ref{fig:neelstate}(a,b) which shows snapshots of the energy
distribution at high and low temperatures. We can clearly observe a
phase transition and the low-temperature phase can be described as a
crystalline N\'eel state expressing up-down energy modulations.
Further, this state is entirely described by energy fluctuations as
there is no sign of long-ranged magnetic order seen in
Fig.~\ref{fig:neelstate}(c,d), as expected.\cite{mishra:207201} A quantity directly probing a crystalline state
as in Fig.~\ref{fig:neelstate}(b) is for example a plaquette structure
factor defined by 
\begin{equation}S_\mathrm{pl}=(1/N)\sum_r^N (-1)^{x_r+y_r}
S_\mathrm{P}(0)S_\mathrm{P}(r),
\end{equation}
where 
\begin{align}\nonumber S_\mathrm{P}(r)&=S_r S_{r+e_x}
+ S_{r+e_y} S_{r+e_x+e_y} + S_{r}S_{r+e_y}\\ &\quad\quad + S_{r+e_x} S_{r+e_x+e_y}
\end{align}
is a four-site spin operator on a plaquette. The sign of $(-1)^{x_r+y_r}$
alternates between A and B plaquettes. We will show below that
$S_\mathrm{pl}$ is indeed an order parameter.
\subsection{Monte Carlo simulations}
\begin{figure}[t]
\centering
\includegraphics[width=0.4\textwidth]{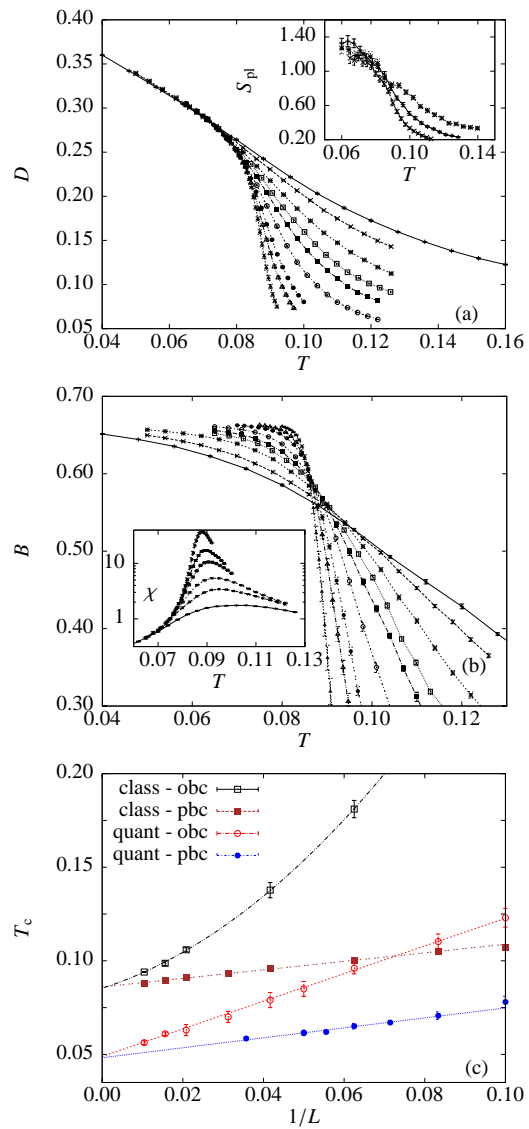}
\caption{\label{fig:results} (color online). Results from classical
  and quantum MC simulations for the 2D  POM. (a)
  The order parameter $D$ for various lattice sizes $L=10$ to $L=96$
  for the classical model employing periodic boundary conditions. The
  inset shows the plaquette structure factor $S_\mathrm{pl}$ for
  $L=20$, $L=32$, and $L=42$. The order parameter indicates a clear
  crystallization effect for $T<0.086$.  (b) The classical Binder
  parameter $B$ close to the transition temperature supporting a
  second-order phase transition. The inset shows the susceptibility
  $\chi$ on a logarithmic scale for $L\geq 16$.  Generally, steeper
  curves in (a) and (b) correspond to larger lattice sizes.  (c) The critical
  temperatures $T_\mathrm{c}$ from FSS of the maxima locations
  of the susceptibility for both the classical and quantum cases and
  two different boundary conditions (open = obc, periodic = pbc).}
\end{figure}%
Having visualized the onset of crystalline order, we now turn to a
more careful discussion of the phase transition from comprehensive
MC runs. Finite lattices are studied using two types of
boundary conditions. In principle, we do not expect unwanted
excitation as in the CM which spoil the FSS
on periodic lattices,\cite{wenzelQCMPRB} but we also study open boundary conditions to
gain further confidence in our results. Open boundary conditions have
the additional advantage that they prefer one sub-lattice over the
other thus possibly stabilizing the ordered phase from the surface. In
the latter case the $|\dots|$ in Eq.~\eqref{eq:orderD} can be also left
away from the definition of the order parameter. In 2D,
simulations were performed of both the classical and quantum cases for various lattice sizes $L=10,\dots,96$, which proved to be sufficiently large. Our analysis to obtain the critical temperatures
is based on $D$, rather than $S_\mathrm{pl}$ because it is easier
measured and is less susceptible to statistical noise. Detection of the
phase boundary proceeds, as usual, by considering the susceptibility
\begin{equation}\chi=N(\langle D^2 \rangle - \langle D \rangle ^2)\,,\end{equation} or the Binder
parameter defined as 
\begin{equation}
B=1- \langle D^4 \rangle /\left( 3 \langle
D^2\rangle^2\right).
\end{equation}%
Figure \ref{fig:results}(a) summarizes data obtained for the classical model
for the order parameter $D$ as well as the structure factor
$S_\mathrm{pl}$ (in inset). Both quantities clearly numerically
establish existence of crystalline order. In Fig.~\ref{fig:results}(b),
data for $B$ and the susceptibility $\chi$ (in inset) is given which
suggests a second-order phase transition at $T_\mathrm{c;cl}=0.0855(4)$
from the crossings of the Binder parameter.  The value of
$B(T_\mathrm{c})=0.610(5)$ at the critical temperature is consistent
with the usual 2D Ising value on the torus topology.\cite{salas-2000-98,selke-2006-51} We further investigate the critical temperature
by a FSS analysis of the maxima of the
susceptibility $\chi$. By fitting the corresponding data in
Fig.~\ref{fig:results}(c) to the usual ansatz
\begin{equation}
  T_{\max}(L)=T_\mathrm{c}+aL^{-1/\nu}+bL^{-w}\,,\end{equation} we obtain
$T_\mathrm{c;cl}=0.0860(2)$ and $T_\mathrm{c;qu}=0.048(1)$ for the
classical and quantum cases, respectively. Here, we assume the 2D Ising exponent $\nu=\nu_{\mathrm{2D Ising}}=1$ (justified by the fit quality and independent fits to the slope of the Binder parameters), and the effective correction term $\sim L^{-w}$ is used
only in fits for open boundary conditions. Since these critical values are
consistently obtained for open and periodic boundary conditions, we
arrive at the important conclusion that FSS in the POM
is well behaved, in clear distinction to the CM.
It is then instructive to compare those values with the related critical
temperatures of the directional-ordering transition in the CM. In
Ref.~\onlinecite{wenzelQCMPRB}, we obtained $T_\mathrm{c;cl}=0.1464(2)$
and $T_\mathrm{c;qu}=0.055(1)$, which leads us to the conclusion that the
geometric variation from Hamiltonian \eqref{eq:hamiltonian0} to
\eqref{eq:hamiltonian1} results in a drastic reduction of $T_\mathrm{c}$ by
$42\%$ for classical fields vs only $13\%$ for the quantum case.
 
In order to obtain further clarity on the type of the transition, we
go one step further and study the model in three dimensions for
various lattice sizes $L=8,\dots,32$. Without showing details but by
performing the same simulations and FSS analysis as before, we obtain
a clear signal for a long-ranged {\it cube-ordered} state below the transition temperatures
$T^\mathrm{3D}_\mathrm{c;cl}=0.365(1)$ and
$T^\mathrm{3D}_\mathrm{c;qu}=0.180(5)$ for classical and quantum
degrees of freedom, respectively. Interestingly, the increase in
$T_\mathrm{c}$ (compared to the 2D transition temperatures) is larger for the classical model.

\begin{figure}
\includegraphics[width=0.9\columnwidth]{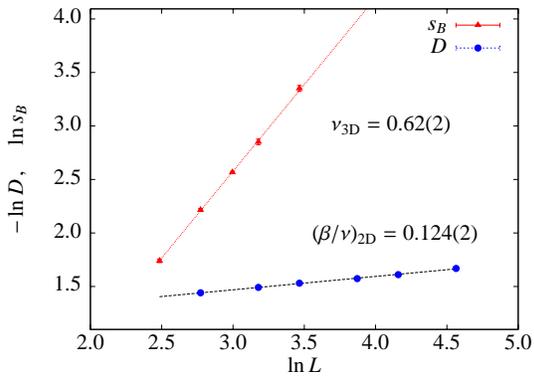}
\caption{\label{fig:critexp} (color online). FSS of the order
  parameter $D$ for a 2D system and for the slope $s_B$ of the Binder
  parameter for a 3D system. Both quantities are calculated at the
  critical temperatures and yield the expected Ising exponents.}
\end{figure}

The proposition that the transition is of Ising type (as suggested by the symmetry and
Binder parameter in Fig.~\ref{fig:results}(b)) should be reflected in
the critical exponents. In Fig.~\ref{fig:critexp} we select two quantities to address this
question, namely the order parameter $D$ for the 2D case and the slope
$s_B$ of the Binder parameter at the critical point for the 3D case.
By performing fits to the ansatz $D\sim L^{-\beta/\nu}$ and $s_B
\sim L^{1/\nu}$, we obtain in the classical cases $(\beta/\nu)_\mathrm{2D} = 0.124(2)$ and
$\nu_\mathrm{3D}=0.62(2)$, which are in excellent agreement with the
theoretical value $(\beta/\nu)_\mathrm{2D}=1/8$ and high-precision
literature\cite{pellisettoreview} on $\nu_\mathrm{3D}$.
Figure~\ref{fig:critexp} shows this scaling versus the lattice size of
both observables at the critical temperatures in a log-log plot.  The
good quality of our data is apparent.  This establishes that the
transition in the POM has Ising exponents for both 2D and 3D, and --
more importantly -- that the {\it energetic} quantity $D$ really
scales like a (magnetic) order parameter. For the quantum case, this
analysis is not so easy but our data is consistent with this
conclusion.

\subsection{Dilution effects in the POM}
\begin{figure}
\centering
\includegraphics[width=0.42\textwidth]{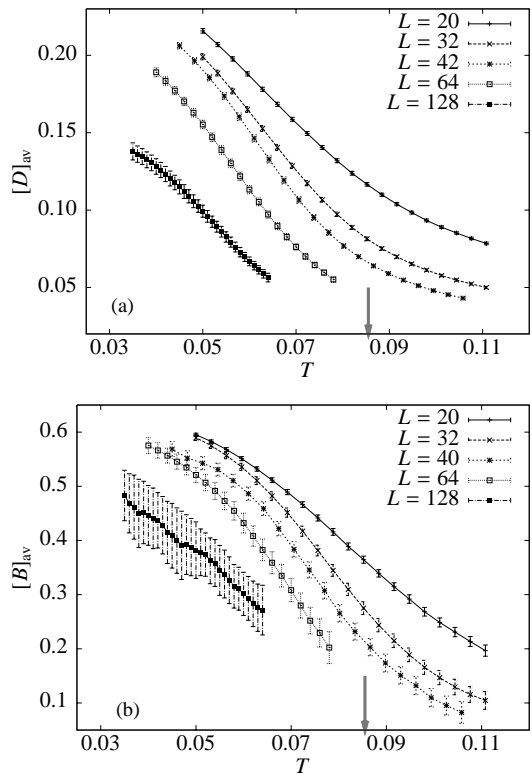}
\caption{\label{fig:disorder} Plot of (a) the disorder averaged (order)
  parameter $[D]_\mathrm{av}$ and (b) the disorder averaged Binder ratio
  $[B]_{\mathrm{av}}$ versus temperature $T$ for $x=1\%$ site dilution in the classical case. The Binder parameter does not show any signature of a remaining phase transition for sizes up to $L=128$ and temperatures down to $T=0.03$. The critical temperature in the clean case is indicated by the arrow.
}
\end{figure}
As the preceding analysis indicates that we have good FSS behavior in
the present model, it is well suited to re-address the important
question of impurity effects\cite{tanaka:256402} in orbital models. To
this end, we employ periodic boundary conditions and define a fraction
$x$ of quenched impurity sites, i.e., we remove each spin with
probability $x$.  Following Ref.~\onlinecite{tanaka:256402}, our
objective is to study the quantity
$g(x)=T_\mathrm{c}(x)/T_\mathrm{c}(0)$ to learn about the degree of
stability of the N\'eel-ordered phase against dilution
disturbances. With $T_\mathrm{c}(x)$ one refers to the critical
temperature of the phase transition with dilution $x$. In order to
access $T_\mathrm{c}(x)$, we performed simulations on lattice sizes
$L=20,32,42,64,128$ (classical case) and $L=20,32$ (quantum case),
where we generated and studied  $100-200$ different disorder
realizations, respectively. The quantities $[D]_\mathrm{av}$,
$[\chi]_\mathrm{av}$, and $[B]_\mathrm{av}$ denote the disorder
averaged values of the respective quantities defined above.

Studying just the (peaks of the) averaged susceptibility or the order
parameter on moderate lattices sizes $(L=20,32,42)$, one could deduce
values for $g(x)$ which are of the order of $g(x)\approx 0.70$ for
the classical case \emph{vs} $g(x)\approx 0.82$ for the quantum model
in case of weak impurity concentration with $x=0.01$. These values seem to be in
qualitative agreement with earlier claims and in support of the
conclusion of Ref.~\onlinecite{tanaka:256402} that quantum
fluctuations make the ordered phase more robust.

However, our simulations on bigger lattices for
$x=0.01$ reveal that the transition is, in fact, vanishing in the
thermodynamic limit, implying $g(x)=0,\;\forall x>0$. This conclusion can be
drawn for example from the data shown in Fig.~\ref{fig:disorder}. While the
order-parameter $[D]_\mathrm{av}$ seems to indicate some crossover which gets weaker for larger $L$, the Binder parameter $[B]_\mathrm{av}$ clearly shows no crossing, i.e., no sign
of critical behavior. Remarkably, quite large lattice sizes $L\approx
128$ are needed to see this. These data obviously rule out the presence
of a true phase transition in the vacancy diluted classical POM.

This situation is further reminiscent of the ordinary 2D Ising model
subjected to a random field (at a fraction $x$ of the sites) which is
known not to exhibit a finite-temperature phase transition from
theoretical\cite{RFIMdc} and numerical works\cite{RFIMdcMC}. While in
the Ising model the random field destroys the $Z_2$ up-down spin
symmetry, dilution in the POM breaks the $A-B$ plaquette
symmetry. Such \emph{random} local symmetry breaking is also known to destroy
long-range order in, e.g., 2D antiferromagnetic Ising models with nearest
and next-nearest neighbor
interactions.\cite{fernandenz:localsymmetrybreaking}

By the same argument it is clear that there is no phase transition in
the 2D classical CM at any finite dilution $x$ and the statements and
conclusions of Ref.~\onlinecite{tanaka:256402} are therefore at
variance with the findings in this work. We also see no argument why
the quantum CM should behave differently in this respect and suspect
that quantum effects merely increase the stability of the low-temperature state on \emph{finite and small} clusters of spins --- an observation that might still
be useful for applications.

\section{Conclusion}
We have introduced and investigated a plaquette orbital model
related to the Kitaev model and the CM. The present work establishes
that this model exhibits an unconventional finite-temperature phase
transition from a disordered to a N\'eel-ordered state in the energy
distribution. It thus displays antiferromagnetic features without
possessing magnetic order. By symmetry arguments and MC simulations,
the critical exponents were determined to be in the Ising
universality class. The geometric variation from the CM to the
plaquette orbital model results in a substantial lowering of the
ordering temperature for the classical model, which is not the case
for quantum spins. 
Our subsequent analysis of the POM in the presence of impurities shows
that long-range ordering is lost for (any) weak disorder
concentration. This finding also sheds concluding light on the
somewhat controversial issue concerning the effect of impurities on the
ordered phase in the compass model.\cite{tanaka:256402,wenzelQCMPRB}
A more detailed analysis of the nature of the ground states for
arbitrary $J_A$, $J_B$ and its quantum phase transitions
\cite{vojta-2003-66,firstorder} would be an interesting continuation
of this work as would be a thorough investigation of the excitations
in the POM. Apart from its possible relevance for protected qubits,
the present model should also be of immediate interest for the physics
of orbital models in relation to transition-metal
oxides.\cite{transmetalScience}

We thank V.~W.~Scarola, B.~Dou\c{c}ot, A.~L\"auchli, and T.~Vojta for discussions. Work supported by the Studienstiftung des deutschen Volkes, the Deutsch-Franz\"osische Hochschule (DFH), the DFG graduate school ``BuildMoNa,'' and NIC J\"ulich.
\bibliographystyle{apsrev}
\bibliography{literature_nourl}

\begin{thebibliography}{21}
\bibitem{footnote1}
For $L=2$ and periodic boundary conditions
  this is not the case as \eqref{eq:hamiltonian1} is equivalent to the
  normal XY model.
\end{thebibliography}
\end{document}